# Free-Electron Qubits

Ori Reinhardt[†], Chen Mechel[†], Morgan Lynch, and Ido Kaminer

Department of Electrical Engineering and Solid State Institute, Technion - Israel Institute of Technology, 32000 Haifa, Israel

[†] *equal contributors*

**Free-electron interactions with laser-driven nanophotonic nearfields can quantize the electrons' energy spectrum and provide control over this quantized degree of freedom. We propose to use such interactions to promote free electrons as carriers of quantum information and find how to create a *qubit* on a free electron. We find how to implement the qubit's non-commutative spin-algebra, then control and measure the qubit state with a universal set of 1-qubit gates. These gates are within the current capabilities of femtosecond-pulsed laser-driven transmission electron microscopy. Pulsed laser driving promise configurability by the laser intensity, polarizability, pulse duration, and arrival times.**

Most platforms for quantum computation today rely on light-matter interactions of bound electrons such as in ion traps [1], superconducting circuits [2], and electron spin systems [3,4]. These form a natural choice for implementing 2-level systems with spin algebra. Instead of using bound electrons for quantum information processing, in this letter we propose using free electrons and manipulating them with femtosecond laser pulses in optical frequencies. Compared to bound electrons, free electron systems enable accessing high energy scales and short time scales. Moreover, they possess quantized degrees of freedom that can take unbounded values, such as orbital angular momentum (OAM), which has also been proposed for information encoding [5-10]. Analogously, photons also have the capability to encode information in their OAM [11,12]. However, implementing even a single qubit, necessitates implementing its (non-commutative) spin algebra, which has so far not been realized in OAM systems. The seemingly obvious choice of implementing qubits on the intrinsic spin of the free electron is not practical because of its weak electromagnetic interaction in optical frequencies [13] (in contrast with lower frequencies, as in electron paramagnetic/spin resonance [14]). Therefore, there remains the nontrivial challenge on the path to creating a laser-driven free-electron qubit: Finding a valid 2-level system capable of implementing spin algebra on the free electron.

We explore a different path for implementing a free-electron qubit. Recent observations of electron-laser interactions provide control over another quantized degree of freedom, the electron energy levels [15-19]. The laser interaction with a free electron creates a quantized ladder of energy levels [15] that provides a synthetic dimension that we will use to encode the qubit state. Experiments manipulating the electron state on this ladder have demonstrated quantum random walks and Rabi oscillations [19], as well as provided a wide variety of methods for coherent control of the electron wavefunction [20,21]. The quantum nature of the interaction has even led to the recent proposal of creating electron-electron entanglement [22,23]. The question at the heart of this paper, of a free-electron qubit, has been a topic of discussion in the community for the last couple of years [24]: Is it possible to utilize electron-laser interactions to imprint a free electron with information in the form of well-defined **qubits?**

To solve this question, and also take it one step further, we follow the DiVincenzo's criteria [25]. Although the electron Hilbert space is unbounded, the task of finding even a single valid qubit is harder than it seems. It is not enough to find two states to define a qubit. The space of states has to be closed under non-commuting quantum operations that must constitute a universal set of 1-qubit gates [3]. These requirements raise the question of whether the available *physical* operations can create a universal set of gates.

Here we utilize the interaction of a free electron with laser pulses to demonstrate a full qubit: A 2-level system implementing a universal set of 1-qubit gates. We show how to algebraically construct the 2-level system that represents the qubit out of the physical electron states, how to implement any quantum gate, and how to measure the qubit state, all using electron–laser interactions in the existing experimental infrastructure of ultrafast transmission electron microscopy (UTEM). Furthermore, we propose the required extensions to having multiple qubits on a single electron.

The most important building-block for our implementation is the free-electron interaction with laser nearfields. This robust interaction forms the basis for photon induced nearfield electron microscopy (PINEM) [15-17]. The PINEM interaction enables the manipulation of the electron quantum state by the laser nearfield in the vicinity of nanophotonic structures or by a field discontinuity on a mirror [21,26] (Fig. 1a). The electron is dressed by the electromagnetic field, forming a bi-directional, infinite ladder of energy states quantized at the photon energy [19] (Fig. 1b). During the interaction, the electron performs a quantum random walk on this energy ladder, and its final wavefunction can be shaped using a general laser pulse, as we have recently shown [27]. Recent publications have shown applications of laser-manipulations of electrons on attosecond timescales [20,28,29]. However, up until now, PINEM has not been used to manipulate quantum information in the form of qubits. In particular, the difficulties of the spin algebra and of defining a valid 2-level system have not been dealt with.

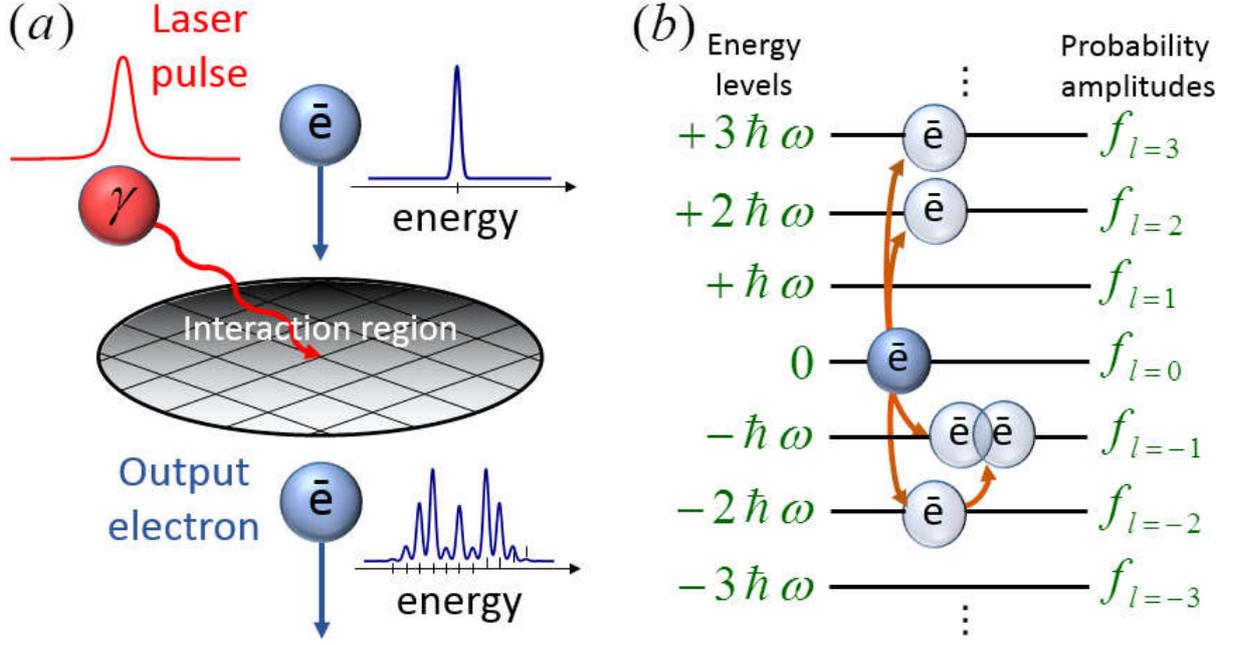

**Figure 1. The platform: Electron-laser interaction**. (a) Interaction between a free electron and an arbitrary laser pulse, which can be demonstrated in the ultrafast transmission electron microscope. The electron wavepacket as a superposition of energy states can be controlled by laser pulses. (b) The electronic ladder states relative to the electron's initial energy. During the interaction, the electron performs a quantum random walk on the energy ladder, where each level is described by its probability amplitude $f_l$.

We first consider the mathematical description of PINEM. An electron with an initial energy $E_0$ interacts with a laser pulse of frequency $\omega$, having a narrow energy spread $\Delta E$ so that $\Delta E < \hbar\omega \ll E_0$. The electron state can be written in the discrete energy basis $|\psi\rangle = \sum_l f_l |l\rangle$, where the integer $l$ represents the number of photons gained/lost by the electromagnetic field during the interaction. This Hilbert space of discrete energies offers a platform to encode information. In this basis, the PINEM interaction operator is represented by an infinite unitary matrix ($U_{PINEM} U_{PINEM}^* = I$):

$$U_{PINEM}(g) = \exp\begin{pmatrix} \cdots & g^* & & & \\ -g & 0 & g^* & & \\ & -g & 0 & g^* & \\ & & -g & 0 & g^* \\ & & & -g & \cdots \end{pmatrix}, \qquad (1)$$

with $g$ the "PINEM field" [16,17,21,30], a dimensionless complex parameter whose magnitude determines the interaction strength between the electron and the laser's electric field (see Supplementary Information part 1). The matrix representation in equation (1) is equivalent to the analytical formula $f_l = e^{il\arg(-g)} J_l(2|g|)$ [16,17,19,21,30], found when applying the PINEM interaction on a single energy state $|0\rangle$ [31].

We can gain further insights from diagonalizing the PINEM operator in equation (1). All eigenvalues lie on the unit circle in the complex plane and are bound azimuthally: $e^{i\phi} \forall |\phi| < 2|g|$, which shows a threshold $g_{th} = \pi/2$ that poses a lower bound on the required laser field amplitude. The corresponding eigenstates are infinite combs of energies, and are thus unphysical, yet expected as the PINEM operation induces quantum walks on the energy ladder that resemble Bloch oscillations, first studied in spatial dimensions [32], and recently in the "synthetic" dimension of the quantized energy levels [33]. The PINEM eigenstates provide a hint for how to construct the 2-level system and resemble Pauli matrix operations.

Interestingly, PINEM alone cannot provide a universal set of qubit gates since different PINEM operators commute regardless of the interaction strength/frequency of the driving laser. Generalizing the interaction by introducing harmonics of the fundamental laser frequency [20,27] (next matrix diagonals in equation (1)) still results in commuting operations. Therefore, an additional component must be included to provide the necessary commutation relations of spin algebra.

To solve this problem, we introduce free-space propagation (FSP) (Fig. 2a), meaning letting the electron propagate freely between laser interactions. Each energy component accumulates phase at a different rate, as was demonstrated inside electron microscopes and other systems [20,28,29]. The FSP operator can be represented as:

$$U_{FSP}(\phi) = \begin{pmatrix} \ddots & & & & & \\ & e^{-4i\phi} & & & & \\ & & e^{-i\phi} & & & \\ & & & 1 & & \\ & & & & e^{-i\phi} & \\ & & & & & e^{-4i\phi} \\ & & & & & & \ddots \end{pmatrix}. \qquad (2)$$

The phase accumulation of each $l$ component along a propagation distance $z$ is given by $\phi_l = 2\pi \frac{z}{z_D} l^2$, with the characteristic length $z_D = 2\beta^2 \gamma^3 \frac{\omega_C}{\omega} \frac{v}{\omega}$, where $\beta = v/c$ is the electron normalized velocity, $\gamma$ the electron Lorentz factor, and $\omega_C$ the Compton frequency (see Supplementary Information part 2).

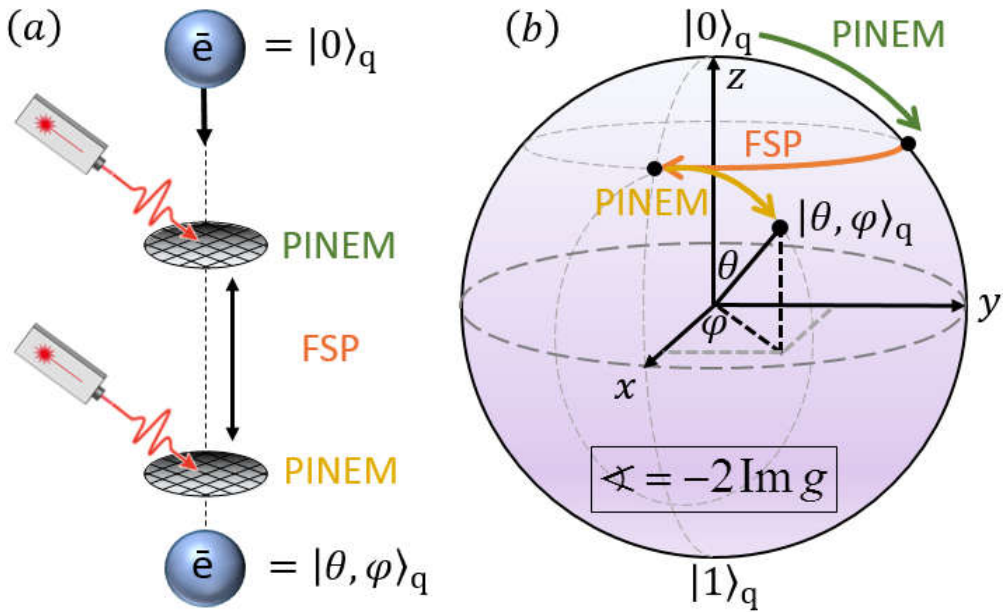

**Figure 2. The quantum gates: Controlling a free-electron qubit**. (a) Illustration of an experimental scheme implementing a quantum gate on a free-electron qubit. A pre-designed quantum computation is translated into gates, mapped to PINEMs and FSPs, and implemented

inside the UTEM. (b) Geometrical representation of the electron qubit and gates on a Bloch-sphere corresponding to the implementation in (a), showing the transformation from the $|0\rangle$ state to an arbitrary point on the Bloch sphere. PINEM operations serve as continuous rotations around the $x$-axis, and FSP operations as $\pi/2$ rotations around the $z$-axis, which together form a universal set of 1-qubit gates.

FSP expands the space of operations we can perform since it acts differently on different energy states and hence does not commute with the PINEM operation. We prove below that the combination of PINEM and FSP enables us to implement a universal set of 1-qubit gates. Interestingly, we note the analogy with the implementations of quantum gates in several other platforms of quantum computing (e.g., coherent control of nuclear magnetic resonance). In many such systems, the quantum gates are implemented by the combination of (radiofrequency/microwave) electromagnetic pulses and proper time delays, corresponding to PINEM and FSP.

Using PINEM and FSP, let us now solve the three challenges of defining the qubit: (1) a 2-level system; (2) a universal set of 1-qubit gates; (3) extraction of its state from the electron wavefunction. For the first challenge, we start from the general electron state $|\psi\rangle = \sum_l \psi_l |l\rangle$ and choose two orthogonal vectors in the following way:

$$\begin{cases} |\psi_0\rangle = (\ldots \quad 0 \quad 1 \quad 0 \quad 1 \quad 0 \quad \ldots)^\dagger \\ |\psi_1\rangle = (\ldots \quad 1 \quad 0 \quad 1 \quad 0 \quad 1 \quad \ldots)^\dagger \end{cases}. \tag{3}$$

We write the electron state as $|\psi\rangle = \alpha |\psi_0\rangle + \beta |\psi_1\rangle + \text{Remainder}$, using a remainder term for the subspace that is orthogonal to $\{|\psi_0\rangle, |\psi_1\rangle\}$. We define the qubit state $|\psi\rangle_q$ by two complex coefficients $\alpha, \beta$, corresponding to the sum of all even/odd energy levels

$$|\psi\rangle_q \equiv \begin{pmatrix} \alpha \\ \beta \end{pmatrix} = \begin{pmatrix} \sum_l \psi_{2l} \\ \sum_l \psi_{2l+1} \end{pmatrix}.$$ Since the remainder is orthogonal to both $|\psi_0\rangle$ and $|\psi_1\rangle$, we can extract the qubit state with a projection $T$:

$$|\psi\rangle_q = T|\psi\rangle \quad, \quad T = |0\rangle_q \langle\psi_0| + |1\rangle_q \langle\psi_1| = \begin{pmatrix} \dots & 1 & 0 & 1 & 0 & 1 & \dots \\ \dots & 0 & 1 & 0 & 1 & 0 & \dots \end{pmatrix}, \tag{4}$$

This projection captures all energy components, increasing its robustness to noise in gate operations or measurements.

It is important to note that as proved below, the Hilbert space spanned by $\{|\psi_0\rangle, |\psi_1\rangle\}$ yields a *valid qubit* because this space is closed under our proposed gates. Furthermore, it can be shown that the norm is conserved $|\alpha|^2 + |\beta|^2 = \text{const}$ by all the interactions, a property originating from the definition of the projection $T$. This means that the qubit gates we apply below are unitary in the qubit space and can be translated to standard Pauli matrices operating on a spin state.

We next turn to prove the second challenge in implementing the qubit: A universal set of 1-qubit gates by using the PINEM and FSP operations in equations (1-2), and applying them on the states in equations (3-4). It can be shown that $|\psi\rangle_q$ is closed under any PINEM operator. For this system to be closed under FSP, we set specific propagation distances, which are integer multiples of $z = z_D/4$. The distance $z_D/4$ adds the phases $\phi_l = \begin{cases} 1 & l \text{ even} \\ i & l \text{ odd} \end{cases}$, forming a $\frac{\pi}{2}$-phase gate, which is exactly the missing ingredient we need for a universal set of 1-qubit gates: It is sufficient to use a (single-frequency) PINEM interaction as in equation (1), and FSP for integer multiples of $z_D/4$.

Figure 2b visualizes the gate operations using the Bloch-sphere representation [3]. For our choice of the qubit states in equations (3-4), PINEM induces a continuous rotation around the $x$-axis, and FSP of distance $z = z_D / 4$ induces a $\pi/2$ rotation around the $z$-axis. Mathematically, these operations in the qubit subspace $|\psi\rangle_q$ take the following forms:

$$U_{FSP} = \begin{pmatrix} 1 & 0 \\ 0 & i \end{pmatrix} = R_z\left(\theta = \frac{\pi}{2}\right) \tag{5a}$$

$$U_{PINEM}(\theta) = \begin{pmatrix} \cos(\theta) & i\sin(\theta) \\ i\sin(\theta) & \cos(\theta) \end{pmatrix} = R_x(\theta = -2\,\mathrm{Im}\,g), \tag{5b}$$

with the rotation angle $\theta \equiv -2\,\mathrm{Im}\{g\}$. Now, using FSP and PINEM together, and noting that $U_{FSP}^{-1} = U_{FSP}^3$, we can implement a $y$-axis rotation using $U_{FSP} U_{PINEM}(\theta) U_{FSP}^3$ with a continuous angle $\theta$. Having continuous rotations around the axes $x$ and $y$, we can implement any 1-qubit gate on the electron with PINEM and FSP (Fig. 2). This set of operations is thus universal.

The third and final challenge required for a free-electron qubit is measuring its state, both amplitudes and phases, meaning a full quantum state tomography [3]. The amplitudes of each electron energy state in the ladder can be obtained using electron energy loss spectroscopy (EELS) that exists in many UTEM systems [15,18,19]. A full construction of the phase of each state was previously demonstrated in the UTEM [20] using a second PINEM operation [34]. Therefore, it is possible to measure the entire qubit state.

The aspect of free-electron qubit state initialization should be discussed as well. Formally, there is no need for such procedure since we have shown universal 1-qubit gates. These gates can be applied on the mono-energetic single energy state $|0\rangle$ (whose projection on the qubit state gives $|0\rangle_q$) to reach any qubit state. The mono-energetic state is produced in the UTEM by electron photo-emission with a low enough energy spread (e.g., [35-37]). Nevertheless, for improved signal to noise ratio (SNR) when calculating the qubit state from electron measurements, it could be valuable to initiate the qubit state with a more complex laser pulse, e.g., made from multiple harmonics of the driving laser [20]. Specifically, in a previous work, we have shown the capability to *shape the electron energy spectrum* [27], which would improve SNR if the electron energy spectrum is shaped to approximate a qubit states from equation (3). This would increase the robustness of our system to errors in gate operations, measurements, and other sources of noise.

Altogether, the experimental realization of our proposal requires PINEM interactions with changing laser field strengths at fixed distances between FSPs. Typical electric fields in the order of $10-100\,\mathrm{MV/m}$ are sufficient for obtaining $|g| \sim \pi/2$ (i.e., any rotation angle), which was already shown in UTEM experiments (e.g., [15,19-21,38,39]). Required FSP lengths, are on the order of $0.1-1\,\mathrm{cm}$, which is also in range of recent experiments [19,20] and in other systems with different spatial constrains [28,29].

Another promising direction for an implementation enables multiple electron-laser interactions in an especially scalable way in multi-pass electron microscopy [40, 41], now being developed for other applications of quantum electron microscopy [42]. This scheme enables the electron to periodically revisit a single interaction point along its trajectory enabling serial PINEM interactions alternating with FSP at the required fixed distances.

Summarizing the requirements, we have several steps for the processing of quantum information with our electrons (Fig. 3). First, we begin from mono-energetic photo-electrons $|0\rangle$ [31] or prepare an initial electron state [27]. Then, we apply desired gates sequentially, where the implementation of each gate using PINEM and FSP is classically calculated in advance. We note that unlike qubit implementations in Floquet systems, here the qubit only experiences periodic laser driving during gate operations, and is a free electron for the rest of the time. We can show that any 1-qubit gate can be implemented by no more than three laser interactions and two propagations. For example, a Hadamard gate can be implemented by $U_{PINEM}(g=\pi)U_{FPS}U_{PINEM}(g=\pi/2)U_{FPS}^3$ and a NOT gate by $U_{PINEM}(g=\pi/2)$ (Fig. 3).

Finally, to measure the final qubit state (amplitude and phase) we find its full wavefunction in the energy basis and project it according to equation (4).

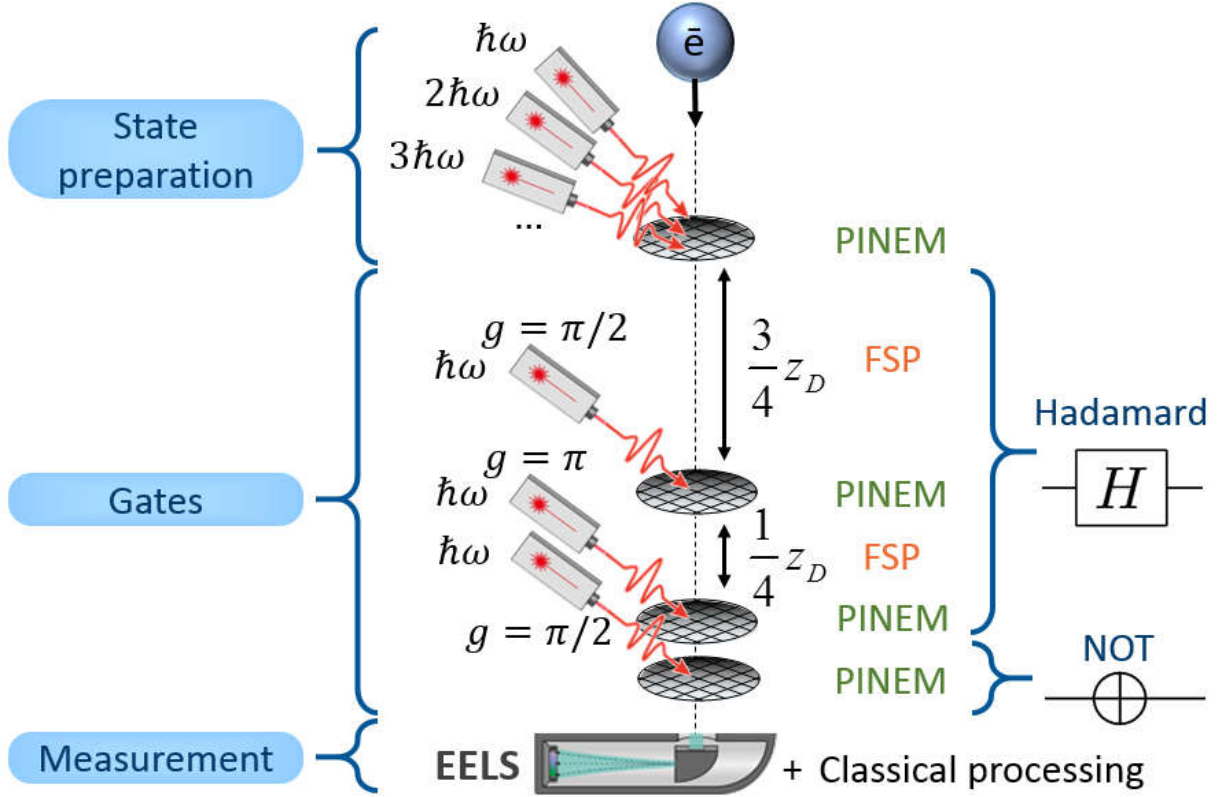

**Figure 3. Satisfying the DiVincenzo criteria: Quantum information processing with free electrons.** Illustration of an experimental scheme. The initial photo-electron is a mono-energetic state $|0\rangle$ or we can perform more advanced energy spectrum shaping with a pre-designed laser pulse [27] for an improved SNR. The next step is applying qubit gates, implemented by PINEM interactions and FSPs. In this specific example, we apply a Hadamard gate followed by a NOT gate. The output electron wavefunction is measured with quantum state tomography [20], before projecting it to retrieve the qubit state.

Looking forward, having a fully controllable free-electron qubit may enable new analytical capabilities in electron microscopes. Future questions include finding how the free-electron qubit state changes by different elastic and inelastic interactions, e.g., with plasmons, excitons, and phonons. It is interesting to explore what information could be extracted from a specimen by using a free-electron qubit as a probe for imaging or energy loss spectroscopy.

It is intriguing to discuss the possible extensions of our scheme to having multiple qubits *on the same single electron.* We suggest generalizing our algebraic method for a larger number of qubits, for example, by using electron ladder states with a periodicity of 4, instead of 2. This definition forms a 4-dimensional Hilbert space upon which 2 qubits can be defined, and so on for larger dimensions that grow exponentially in the number of qubits. A remaining challenge is proving that the available set of quantum operators for two qubits is universal, similarly to the procedure we have shown for the case of 1-qubit gates. For that, we note that PINEM and FSP has an additional algebraic richness than what we have exploited in this work, including the possibility of having multi-frequency PINEM.

Furthermore, the free-electron qubit can be directly extended to other types of quantum units of information, such as qutrits. This method can be understood as "Hilbert space engineering" – taking an immensely large electron energy space and finding a smaller synthetic dimension in it, tailor-made for a certain task.

Importantly, we have recently shown the ability to occupy ~1000 energy levels [43] using a single-frequency PINEM, and we expect up to about $\log_2 1000 \approx 10$ qubits to be possible on a single electron using the existing capabilities of UTEM systems.

Another direction to implement multiple qubits on the same single electron is by having non-overlapping energy ladders on the electron, each representing a different qubit. Each ladder would have the same unit-spacing in energy (i.e. $\hbar\omega$), and the shift between ladders would be a fraction of that spacing. In this case, the algebra is further enriched by choosing different propagation distances and PINEM frequencies. This method is limited by the electron energy spread and spectrometer resolution, and the available operations have to be proven to be universal, which we leave for future work.

Although speculative at this point, it is worth envisioning mechanisms of electron-electron interactions to create connectivity between different qubit groups on different electrons. Such mechanisms could involve strong coupling to a photonic cavity [39] that helps communicate between electrons and entangle them [22,23] to enable a larger computational space.

Note that ideas of implementing qubit algebra in a synthetic Hilbert space were investigated with other infinite-dimensional continuous-variables degrees of freedom [44], such as in time and energy domains by photons [45,46] or within molecular qudits [47]. Our work presents a conceptually different physical system, developing the building blocks of quantum computation on a single free electron for the first time. One strong advantage of our proposal is configurability: changing the electron Hilbert space and the quantum gates is a matter of changing laser pulse properties, unlike the case in fabricated photonic waveguide arrays or superconducting circuits.

To conclude, we propose the idea of manipulating quantum information on a free electron by "algebraically engineering" the electron's Hilbert space into a qubit. We propose an implementation using electron-laser interactions in the UTEM and prove that our method defines a valid qubit with a universal set of gates. The conceptual ideas we develop take a step toward **electron microscopes as platforms of quantum information processing**. Establishing a framework for manipulating multiple qubits on a single free electron suggests the possibility of having quantum computation on qubits that are not *physically separated*, but only *"algebraically" separated*.